\documentclass[aps,pra,superscriptaddress,twocolumn,showpacs]{revtex4}

\renewcommand*{\[}{\begin{equation}}

\renewcommand*{\]}{\end{equation}}

\newcommand{\myscaleboxa}[1]{\scalebox{0.5}[0.5]{#1}}
\newcommand{\myscaleboxb}[1]{\scalebox{0.8}[0.8]{#1}}
\newcommand{\myscaleboxc}[1]{\scalebox{0.35}[0.35]{#1}}

\usepackage{epsfig}

\usepackage{hyperref}

\begin{document}

\title{Medium propagation effects in high harmonic generation of Ar and N$_{2}$}

\author{Cheng Jin}

\affiliation{J. R. Macdonald Laboratory, Physics Department, Kansas
State University, Manhattan, Kansas 66506-2604, USA}

\author{Anh-Thu Le}
\affiliation{J. R. Macdonald Laboratory, Physics Department, Kansas
State University, Manhattan, Kansas 66506-2604, USA}

\author{C. D. Lin}
\affiliation{J. R. Macdonald Laboratory, Physics Department, Kansas
State University, Manhattan, Kansas 66506-2604, USA}

\date{\today}

\begin{abstract}
We report theoretical calculations of high harmonic generation (HHG)
by intense infrared lasers in atomic and molecular targets taking
into account the macroscopic propagation of both fundamental and
harmonic fields. On the examples of Ar and N$_2$, we demonstrate
that these {\it ab initio} calculations are capable of accurately
reproducing available experimental results with isotropic and
aligned target media. We further present detailed analysis of HHG
intensity and phase, under various experimental conditions, in
particular, as the wavelength of the driving laser changes. Most
importantly, our results strongly support the factorization of HHG
at the macroscopic level into a product of a returning electron wave
packet and the photorecombination transition dipole, under typical
experimental conditions. This implies that the single-atom/molecule
structure information can be retrieved from experimentally measured
HHG spectra.

\end{abstract}


\maketitle

\section{Introduction}
High harmonic generation (HHG) is an extreme nonlinear optical
process in which an intense ultrafast infrared light is efficiently
converted to an ultrafast coherent extreme ultraviolet (XUV) or soft
X-ray light. In the last two decades, HHG has been widely studied
for its potential as a short-wavelength light source, either in the
form of a useful bright, coherent tabletop light down to the
water-window region (280-540 eV) \cite{andy-Sci-98,randy-Sci-02}, or
the production of ultrashort light pulses such as single attosecond
pulses and attosecond pulse trains
\cite{krausz-rmp-09,agostini-rpp-04}. Recently, HHG itself has also
been shown to have the potential to image molecular structure with
sub-Angstrom precision in space and sub-femtosecond precision in
time
\cite{itatani-nature-04,olga-nature-09,hans-nature-10,hoang-pra-07}.
The basic principle of harmonic emission in a gas medium is well
understood qualitatively. When an atom or molecule is exposed to an
intense laser field, first, at a certain time an electron wave
packet tunnels through the barrier formed by the combined atomic and
laser fields. Next, it propagates in the laser field and may be
driven back to recollide with the target ion. High harmonics are
generated when the returning electrons recombine with the ion, and
convert the energy gained in the laser field to high-energy photons
\cite{krausz-prl-92,corkum-prl-93}. Since the laser field interacts
with a macroscopic medium, and the harmonics from all atoms or
molecules are generated coherently, a full description of the
experimentally observed HHG spectra requires the treatment of the
nonlinear propagation of the fundamental laser beam together with
the harmonics in the medium. Thus the study of HHG consists of two
parts. The first one is the calculation of the induced dipole by
each atom or molecule in the laser field. This is to be carried out
quantum mechanically by solving the time-dependent Schr\"odinger
equation (TDSE) or equivalents accurately. The second one is to
solve the nonlinear propagation of the fundamental laser pulse and
the harmonic fields in the medium by using Maxwell's equations. As
in all nonlinear processes, an efficient harmonic generation
requires good phase matching from all the elementary induced
dipoles, which in turn depends on laser's properties like intensity,
pulse duration,  pulse shape, and target properties such as gas
pressure, position of gas jet with respect to the laser focus, in
addition to the linear and nonlinear responses of gas in the light
fields. Clearly, a full understanding of HHG cannot be reached until
all of these effects are properly described theoretically. This is
especially important if one is to use HHG to image the structure of
molecules. Since HHG spectra are sensitive to the detailed
experimental conditions which usually cannot be accurately
determined in a given experiment, how to extract useful quantitative
structure information of individual target molecules in the gas
medium is clearly a challenge.

The most accurate way to obtain the induced dipole of an atom or
molecule in a laser field is to solve the TDSE numerically. Since
this approach is quite time consuming and the calculations have to
be carried out for hundreds of laser peak intensities in order to
describe the nonuniform laser distributions inside a focused laser
beam, this is rarely done in existing studies  including macroscopic
propagation effect of HHG \cite{mett-pra-06}. Instead, the much
simpler strong field approximation (SFA), or the so-called
Lewenstein model \cite{Lewen}, is often used to calculate the single
atom response. Despite of this limitation, the temporal and spatial
properties of HHG observed experimentally have been reasonably
understood from such SFA-based calculations. On the other hand, in a
few examples, macroscopic HHG spectra obtained using TSDE-calculated
induced dipoles do show significant quantitative discrepancy
compared to SFA-based calculations~\cite{mett-pra-02,Mette-jpb}, and
such studies have been limited to a few atomic gases only. To image
the structure of molecules from the experimental HHG spectra, one
first needs to be able to describe HHG spectra from molecules
including macroscopic effects.

In this article, we demonstrate an accurate and efficient method for
calculating the HHG spectra from a macroscopic atomic or molecular
gaseous medium. The method is based on our recently developed
quantitative rescattering (QRS) theory
\cite{lin-jpb-10,toru-2008,at-2009} which allows us to calculate the
induced dipole of an atom or molecule in a laser field with accuracy
comparable to those obtained from solving TDSE, yet with computing
time comparable to those by using the SFA. More importantly,
according to QRS, one can express the complex induced dipole moment
as the product of a complex returning wave packet and a complex
photorecombination (PR) transition dipole moment, where the former
depends on the properties of the laser and the latter is solely the
property of the target. In fact, the elementary PR transition dipole
moment is identical to the laser-free elementary photoionization
(PI) transition dipole moment which has been well studied in the
last few decades. Using the QRS, we further show that the complex
returning wave packet can be obtained from the SFA. The validity of
the QRS, at the single atom or single molecule level, has been
carefully calibrated against TDSE results for one-electron model
atoms, and against experimental HHG spectra from molecules
\cite{lin-jpb-10}. Clearly such comparison is incomplete without
considering the macroscopic propagation effects. In an earlier
paper, based on the laser induced dipoles calculated using the QRS,
Jin {\it et al}. \cite {jin-2009} studied the macroscopic
propagation effects of the HHG of rare gases theoretically for the
situation where the laser intensity and the gas pressure are small.
Under this limit, the fundamental laser field is assumed   not
modified during the propagation. It was shown that the macroscopic
HHG spectra after propagation can be expressed as the product of a
``macroscopic wave packet" and the same single-atom PR transition
dipole. This result is very important since it enables us to extract
target structure from the experimentally measured HHG spectra, thus
paving the way for using infrared lasers for time-resolved imaging
of transient molecules. In the present article, we extend the work
of Jin {\it et al}. \cite{jin-2009} to higher laser intensities and
gas pressures at which the nonlinear propagation of the fundamental
field is considered. We then examine the theoretically simulated HHG
spectra of Ar and compare them directly with experimental data. We
further extend the method to include molecular targets, which are
aligned or isotropically distributed.

In Section II, we summarize the method and the essential equations
for describing the macroscopic propagation, and the calculation of
single-atom or single-molecule induced dipoles in the QRS model. We
also stress that the HHG spectra should be calculated for each
specific experimental condition. Within the QRS, we can define a
``macroscopic wave packet" (MWP) which will reflect the effect of
lasers and the consequence of propagation in the medium. In Section
III, the results are shown and analyzed. First, we consider HHG
spectra of Ar generated by 1200 nm lasers, and show that the
experimental HHG spectra from 30-90 eV can be accurately reproduced
theoretically based on the QRS, but not on the commonly used SFA. In
fact, the HHG spectra depend on the position where and how the XUV
light is measured. We show how the two-dimensional HHG spectra,
their global behavior and individual single harmonics in the far
field, depend on the gas pressure and the pulse length. We also
study the spatial distributions of individual harmonics in the near
field and in the far field.  The phase and the amplitude of the
harmonics after propagation are also analyzed, for harmonics along
the propagation axis and off the axis. The harmonics are found to be
always varied with these parameters and the different experimental
conditions. However, we find that all the differences can be
attributed to the different MWP's. Thus the dependence of MWP on the
gas-jet position with respect to the laser focus, the degree of
phase matching with respect to the gas pressure for individual
harmonics are investigated. Since phase-matching condition is also
dependent on the wavelength of the laser used, we investigate how
macroscopic HHG scales with the laser wavelength. The QRS has been
used to obtain induced dipoles from molecules, so we extend the
present work to molecular targets. We consider the isotropic and
partially aligned molecules, and demonstrate that the experimental
HHG spectra of N$_2$ molecules from recent measurements and the
present calculations are in good agreement. In Section IV, we
summarize and discuss future perspective before concluding this
paper.

\section{Theoretical method}
\subsection{Propagation of the fundamental field}
In an ionizing gas, the propagation of a fundamental driving laser
is affected by refraction, nonlinear self-focusing, ionization, and
plasma defocusing. The pulse evolution in such a medium is described
by a three-dimensional (3-D) Maxwell's wave equation
\cite{IEEE-1997,Tosa-pra-2003,Geissler-prl-99}:
\begin{eqnarray}
\label{fund-time}\nabla^{2}E_{1}(r,z,t)&&-\frac{1}{c^{2}}\frac{\partial^{2}E_{1}(r,z,t)}{\partial
t^{2}}= \mu_{0}\frac{\partial J_{\text {abs}}(r,z,t)}{\partial
t}\nonumber\\&&+\frac{\omega_{0}^{2}}{c^{2}}(1-\eta_{\text
{eff}}^{2})E_{1}(r,z,t),
\end{eqnarray}
where $E_{1}(r,z,t)$ is the transverse electric field of the fundamental laser
pulse with frequency $\omega_{0}$. In cylindrical coordinates,
$\nabla^{2}=\nabla_{\bot}^{2}+\partial^{2}/\partial z^{2}$,  where $z$ is the
axial propagation direction. The effective refractive index of the gas medium
can be written as
\begin{eqnarray}
\label{eff}\eta_{\text
{eff}}(r,z,t)=\eta_{0}(r,z,t)+\eta_{2}I(r,z,t)-\frac{\omega_{\text
p}^{2}(r,z,t)}{2\omega_{0}^{2}}.
\end{eqnarray}
The first term $\eta_{0}=1+\delta_{1}-i\beta_{1}$ takes into account of
refraction ($\delta_{1}$) and absorption ($\beta_{1}$) effects of the neutral
atoms, the second term accounts for the optical Kerr nonlinearity which depends
on laser intensity $I(t)$, and the third term is from free electrons which
contains the plasma frequency $\omega_{\text p}=[e^{2}n_{\text
e}(t)/(\varepsilon_{0}m_{\text e})]^{1/2}$, where $m_{\text e}$ and $e$ are the
mass and charge of an electron, respectively, and $n_{\text e}(t)$ is the
density of free electrons. The absorption term $J_{\text {abs}}(t)$ due to the
ionization of the medium is expressed as \cite{Mette-jpb,Rae-pra-1992}
\begin{eqnarray}
J_{\text {abs}}(t)=\frac{\gamma(t)n_{\text e}(t)I_{\text
p}E_{1}(t)}{|E_{1}(t)|^{2}},
\end{eqnarray}
where $\gamma(t)$ is the ionization rate, and $I_{\text p}$ is the ionization
potential. This term is usually small under the conditions for harmonic
generation \cite{Mette-jpb,Rae-pra-1992}.

The absorption effect ($\beta_{1}$) on the fundamental laser field
caused by neutral atoms is in general small, so it is neglected. We
only keep the real terms in the refractive index $\eta_{\text
{eff}}$, and Eq.~(\ref{fund-time}) can be written as
\begin{eqnarray}
&&\nabla^{2}E_{1}(r,z,t)-\frac{1}{c^{2}}\frac{\partial^{2}E_{1}(r,z,t)}{\partial
t^{2}}=\mu_{0}\frac{\partial J_{\text {abs}}(r,z,t)}{\partial t}
\nonumber\\&&+\frac{\omega_{\text
p}^{2}}{c^{2}}E_{1}(r,z,t)-2\frac{\omega_{0}^{2}}{c^{2}}(\delta_{1}+\eta_{2}I)E_{1}(r,z,t).
\end{eqnarray}
By going to a moving coordinate frame ($z^{\prime}=z$ and
$t^{\prime}=t-z/c$) and employing the paraxial approximation (i.e.,
neglecting $\partial^{2}E_{1}/\partial z^{\prime 2}$), we obtain
\cite{Priori-pra-2000}
\begin{eqnarray}
\label{fund-moving}&&\nabla_{\bot}^{2}E_{1}(r,z',t')-\frac{2}{c}\frac{\partial^{2}E_{1}(r,z',t')}{\partial
z'\partial t'}=\mu_{0}\frac{\partial J_{\text {abs}}(r,z',t')}{\partial t'}
\nonumber\\&&+\frac{\omega_{\text p}^{2}}{c^{2}}E_{1}(r,z',t')
-2\frac{\omega_{0}^{2}}{c^{2}}(\delta_{1}+\eta_{2}I)E_{1}(r,z',t').
\end{eqnarray}
The temporal derivative in Eq.~(\ref{fund-moving}) can be eliminated
by a Fourier transform, yielding the equation
\begin{eqnarray}
\label{fund-freq}\nabla_{\bot}^{2}\tilde{E}_{1}(r,z',\omega)-\frac{2i\omega}{c}\frac{\partial
\tilde{E}_{1}(r,z',\omega)}{\partial z'}=\tilde{G}(r,z',\omega),
\end{eqnarray}
where
\begin{eqnarray}
\tilde{E}_{1}(r,z',\omega)=\hat{F}[E_{1}(r,z',t')],
\end{eqnarray}
and
\begin{eqnarray}
\tilde{G}(r,z',\omega)&&=\hat{F}\{\mu_{0}\frac{\partial J_{\text
{abs}}(r,z',t')}{\partial t'}+\frac{\omega_{\text
p}^{2}}{c^{2}}E_{1}(r,z',t')
\nonumber\\&&-2\frac{\omega_{0}^{2}}{c^{2}}[\delta_{1}
+\eta_{2}I(r,z',t')]E_{1}(r,z',t')\},
\end{eqnarray}
where $\hat{F}$ is the Fourier transform operator acting on the
temporal coordinate.

The plasma frequency $\omega_{\text p}(r,z',t')$ is determined by
the free-electron density $n_{\text e}(t')$, and $n_{\text e}(t')$
can be calculated as following
\begin{eqnarray}
\label{free-electron}n_{\text
e}(r,z',t')=n_{0}\{1-\exp[-\int_{-\infty}^{t'}\gamma(r,z',\tau)d\tau]\},
\end{eqnarray}
where $n_{0}$ is the neutral atom density, and $\gamma(r,z',\tau)$
is the ionization rate calculated from Ammosov-Delone-Krainov (ADK)
theory \cite{adk,tong-pra-2002,tong-jpb}. The refraction coefficient
$\delta_{1}$, depending on the pressure and temperature of the gas
medium, is obtained from the Sellmeier equation
\cite{ref-2008,ref-1974}. The second-order refractive index
$\eta_{2}$, also depending on pressure of the gas medium, can be
calculated through third-order susceptibility $\chi^{(3)}$, which
can be measured from experiments \cite{Opt-1985,Li-pra-1989}. Note
that the relationship between $\eta_{2}$ and $\chi^{(3)}$ in Koga
{\it et al}. \cite{Koga-2000} differs from that in Boyd \cite{Boyd}
since the latter is derived by using time-averaged intensity of the
optical field. The fundamental laser field is assumed to be Gaussian
both in space and in time at the entrance of a gas jet ($z'=z_{\text
{in}}$), and the pressure is assumed constant within the gas jet.

\subsection{Propagation of the harmonic field}
The 3-D propagation equation of the harmonic field is described by
\cite{Priori-pra-2000,Mette-jpb,tosa-pra-2005}
\begin{eqnarray}
\label{harm-time}\nabla^{2}E_{\text
h}(r,z,t)-\frac{1}{c^{2}}\frac{\partial^{2}E_{\text
h}(r,z,t)}{\partial
t^{2}}=\mu_{0}\frac{\partial^{2}P(r,z,t)}{\partial t^{2}},
\end{eqnarray}
where $P(r,z,t)$ is the polarization depending upon the applied
optical field $E_{1}(r,z,t)$. In this equation, the free-electron
dispersion is neglected because the frequencies of high harmonics
are much higher than the plasma frequency. Again going to a moving
coordinate frame and using the paraxial approximation,
Eq.~(\ref{harm-time}) becomes
\begin{eqnarray}
\label{harm-moving}\nabla_{\bot}^{2}E_{\text
h}(r,z',t')-\frac{2}{c}\frac{\partial^{2}E_{\text
h}(r,z',t')}{\partial z'\partial
t'}=\mu_{0}\frac{\partial^{2}P(r,z',t')}{\partial
t'^{2}}.&&\nonumber\\&&
\end{eqnarray}
We eliminate the temporal derivative by a Fourier transform, obtaining the equation
\begin{eqnarray}
\label{harm-freq}\nabla_{\bot}^{2}\tilde{E}_{\text
h}(r,z',\omega)-\frac{2i\omega}{c}\frac{\partial \tilde{E}_{\text
h}(r,z',\omega)}{\partial
z'}=-\omega^{2}\mu_{0}\tilde{P}(r,z',\omega),&& \nonumber\\&&
\end{eqnarray}
where
\begin{eqnarray}
\tilde{E}_{\text h}(r,z',\omega)=\hat{F}[E_{\text h}(r,z',t')],
\end{eqnarray}
and
\begin{eqnarray}
\tilde{P}(r,z',\omega)=\hat{F}[P(r,z',t')].
\end{eqnarray}
The source term on the right-hand side of Eq.~(\ref{harm-freq})
describes the response of the medium to the laser field and includes
both linear and nonlinear terms. It is convenient to separate the
polarization field into  linear and nonlinear components
$\tilde{P}(r,z',\omega)=\chi^{(1)}(\omega)\tilde{E}_{\text
h}(r,z',\omega)+\tilde{P}_{\text {nl}}(r,z',\omega)$, where the
linear susceptibility $\chi^{(1)}(\omega)$ includes both linear
dispersion and absorption through its real and imaginary parts,
respectively. The nonlinear polarization term $\tilde{P}_{\text
{nl}}(r,z',\omega)$ can be expressed as
\begin{eqnarray}
\label{pola}\tilde{P}_{\text
{nl}}(r,z',\omega)=\hat{F}\{[n_{0}-n_{\text
e}(r,z',t')]D(r,z',t')\},
\end{eqnarray}
where $n_{\text e}(r,z',t')$ is calculated from
Eq.~(\ref{free-electron}), and $D(r,z',t')$ is the single-atom
induced dipole moment caused by the fundamental driving laser field.

The refractive index
$n(\omega)=\sqrt{1+\chi^{(1)}(\omega)/\varepsilon_{0}}$ \cite{Boyd}
is related to atomic scattering factors by
\begin{eqnarray}
n(\omega)&&=1-\delta_{\text h}(\omega)-i\beta_{\text h}(\omega)
\nonumber\\&& =1-\frac{1}{2\pi}n_{0}r_{0}\lambda^{2}(f_{1}+if_{2}),
\end{eqnarray}
where $r_{0}$ is the classical electron radius, $\lambda$ is the
wavelength, $n_{0}$ is again the neutral atom density, and $f_{1}$
and $f_{2}$ are atomic scattering factors which can be obtained from
Refs. \cite{nist,Henke}. Note that $\delta_{\text h}(\omega)$ and
$\beta_{\text h}(\omega)$ account for the dispersion and absorption
of the medium on the harmonics, respectively. Finally
Eq.~(\ref{harm-freq}) can be written as
\begin{eqnarray}
\label{harm-final}&&\nabla_{\bot}^{2}\tilde{E}_{\text
h}(r,z',\omega)-\frac{2i\omega}{c}\frac{\partial \tilde{E}_{\text
h}(r,z',\omega)}{\partial z'}
\nonumber\\&&-\frac{2\omega^{2}}{c^{2}}(\delta_{\text
h}+i\beta_{\text h})\tilde{E}_{\text h}(r,z',\omega)
=-\omega^{2}\mu_{0}\tilde{P}_{\text {nl}}(r,z',\omega),\nonumber\\
\end{eqnarray}
where the nonlinear polarization as the source of the harmonics is
explicitly given. After the propagation in the medium, we obtain the
near-field harmonics at the exit face of the gas jet ($z'=z_{\text
{out}}$).

As presented in Ref. \cite{jin-2009}, when both the pressure and
laser intensity are low, the fundamental field is not modified
through the medium. In other words, the source term in
Eq.~(\ref{fund-time}) can be taken as zero. Then the fundamental
laser field, assuming to be a Gaussian beam spatially, is given
approximately in an analytical form. For the harmonic field, the
dispersion and absorption effects from the medium, which are
explicitly expressed as a dispersion-absorption term in
Eq.~(\ref{harm-final}) are not included. These effects would become
important if the gas pressure is high. For molecular targets, we
will limit ourselves to experiments carried out under the conditions
of low laser intensity and low gas pressure. Note that Eqs.
(\ref{fund-freq}) and (\ref{harm-final}) are solved using a
Crank-Nicholson routine for each value of $\omega$. Typical
parameters used in the calculations are 200$\sim$300 grid points
along the radial direction and 400 grid points along the
longitudinal direction.

\subsection{Far-field harmonic emission}
Experimentally, harmonics are not measured at the exit face of a gas
medium. They may go through a slit, an iris or a pinhole, or be
reflected by a mirror before they reach the detector. The far-field
harmonic emissions can be obtained from near-field harmonic
emissions at the exit face of a gas medium through a Hankel
transformation \cite{far-field,L'Huillier-1992,tosa-2009}
\begin{eqnarray}
\label{far-hhg}{E}_{\text h}^{\text f}(r_{\text f},z_{\text
f},\omega)=&&ik\int\frac{\tilde{E}_{\text h}(r,z',\omega)}{z_{\text
f}-z'}J_{0}(\frac{k rr_{\text f}}{z_{\text
f}-z'})\nonumber\\&&\times \exp[\frac{i k(r^{2}+r_{\text
f}^{2})}{2(z_{\text f}-z')}] r dr,
\end{eqnarray}
where $J_{0}$ is the zero-order Bessel function, $z_{\text f}$ is
the far-field position from the laser focus, $r_{\text f}$ is the
transverse coordinate in the far field, and the wave vector $k$ is
given by $k=\omega/c$.

Suppose the harmonics in the far field are collected from an
extended area, the power spectrum of the macroscopic harmonics is
obtained by integrating harmonic yields over the area:
\begin{eqnarray}
\label{total-hhg}S_{\text h}(\omega)\propto\int\int|{E}_{\text
h}^{\text f}(x_{\text f},y_{\text f},z_{\text
f},\omega)|^{2}dx_{\text f}dy_{\text f},
\end{eqnarray}
where $x_{\text f}$ and $y_{\text f}$ are the Cartesian coordinates
on the plane perpendicular to the propagation direction, and
$r_{\text f}=\sqrt{x_{\text f}^{2}+y_{\text f}^{2}}$.

\subsection{Quantitative rescattering (QRS) theory}
In this work, the single-atom (or single-molecule) induced dipole
moment $D(t')$ in Eq.~(\ref{pola}) is calculated quantum
mechanically using the QRS theory. A detailed discussion of QRS for
HHG from atoms or molecules is given in Ref. \cite{at-2009}. We
briefly discuss the QRS theory for atoms and molecules separately in
the following.

\subsubsection{Atomic target}
According to the QRS, the induced dipole moment $D(\omega)$ can be
written as \cite{at-2008}
\begin{eqnarray}
\label{qrs-phase}D(\omega)=W(\omega)d(\omega),
\end{eqnarray}
where $d(\omega)$ is the complex photorecombination (PR) transition
dipole matrix element, and $W(\omega)$ is the complex microscopic
wave packet. $|W(\omega)|^{2}$ describes the flux of the returning
electrons and is the property of the laser only. The QRS is a simple
model that improves the SFA. It replaces the plane wave used in the
SFA with accurate  scattering wave in the calculation of PR
transition dipole matrix elements, while the returning microscopic
wave packet is the same as that in the SFA. Since the electron wave
packet after tunneling but before the recombination is governed
mostly by the laser field while the electron is far away from the
target ion, and such interaction is fully described by the SFA, this
explains why the electron wave packet (its dependence on the
momentum of returning electrons) derived from the SFA is accurate.
In practical applications, the QRS obtains the induced dipole moment
by
\begin{eqnarray}
\label{qrs}D^{\text {QRS}}(\omega)=D^{\text
{SFA}}(\omega)\frac{d^{\text {QRS}}(\omega)}{d^{\text
{SFA}}(\omega)},
\end{eqnarray}
where both $D^{\text {SFA}}(\omega)$ and $d^{\text {QRS}}(\omega)$
are complex numbers, while $d^{\text {SFA}}(\omega)$ is either a
pure real or pure imaginary number. Within the single active
electron (SAE) approximation, we calculate $d^{\text {QRS}}(\omega)$
using ``exact" numerical wave functions for the bound and continuum
states. For Ar, we use the model potential given by M\"uller
\cite{Muller},
\begin{eqnarray}
\label{muller}V(r)=-[1+Ae^{-r}+(17-A)e^{-Cr}]/r,
\end{eqnarray}
with(A=5.4, C=3.682). In this model, spin-orbit interaction is
neglected. The parameters have been chosen such that the minimum in
the photoionization (or photorecombination) cross section is
reproduced correctly. We comment that in principle the parameters in
Eq.~(\ref{qrs}) can be generalized to many-electron wave functions
if needed.

\subsubsection{Molecular target}
Within the QRS theory, the induced dipole moment $D(\omega,\theta)$
for a fixed-in-space molecule is given explicitly by
\begin{eqnarray}
D(\omega,\theta)=N(\theta)^{1/2}W(\omega)d(\omega,\theta),
\end{eqnarray}
where $N(\theta)$ is the alignment-dependent ionization probability,
$W(\omega)$ is the microscopic wave packet, and $d(\omega,\theta)$
is the alignment-dependent transition dipole (complex in general).
Here $\theta$ is angle between the molecular axis with respect to
the laser's polarization. We limit ourselves here to linearly
polarized lights, linear molecules and consider the parallel
component of HHG with respect to the laser polarization only. Thus
only the parallel component of the transition dipole
$d(\omega,\theta)$ is needed in the calculation. Note that
$W(\omega)$ does not depend on the alignment angle $\theta$, and it
can be calculated formally as
\begin{eqnarray}
\label{mol-qrs}W(\omega)=\frac{D(\omega,\theta)}{N(\theta)^{1/2}d(\omega,\theta)}.
\end{eqnarray}
Recall that $W(\omega)$ can be obtained from SFA, where all the
matrix elements above are calculated by replacing the continuum
waves by plane waves. Since the wave packet $W(\omega)$ is
independent of the alignment angle  $\theta$, it needs to be
calculated only once  for a given angle $\theta$. In the QRS, the
single-molecule induced dipole moment by the same laser is then
obtained from Eq.~(\ref{mol-qrs}) by combining with the accurate
$d(\omega,\theta)$ obtained from quantum chemistry code
\cite{Lucchese-pra-82} and with the tunneling ionization rate
$N(\theta)$ obtained from the MO-ADK theory
\cite{tong-pra-2002,zhao-pra-10}. Applications of the QRS for HHG
from single molecules have been investigated previously
\cite{at-2009,at-prl-2009,at-jpb-2009}.

Linear molecules can only be  partially aligned when they are placed
in a short laser field (pump laser) \cite{jin-pra-10}. The intensity
of the aligning laser is usually weak and not tightly focused such
that it can be assumed to be constant within the gas medium. In
other words, the degree of molecular alignment is not varied in the
medium. In the following, the polarization of aligning laser is
assumed to be parallel to the probe laser. The averaged induced
dipole from the partially aligned molecules at each point in the gas
medium is then obtained by coherently averaging induced dipole
moments over the molecular angular distributions, i.e.,
\begin{eqnarray}
\label{avg-in-dip}D^{\text
{avg}}(\omega)=\int^{\pi}_{0}D(\omega,\theta)\rho(\theta)\sin\theta
d\theta,
\end{eqnarray}
where $\rho(\theta)$ is the angular (or alignment) distribution of
the molecules. Similarly, the free electron density in
Eq.~(\ref{pola}) is replaced by the averaged one:
\begin{eqnarray}
n_{\text e}^{\text {avg}}(t')=\int^{\pi}_{0}n_{\text
e}(t',\theta)\rho(\theta)\sin\theta d\theta,
\end{eqnarray}
where $n_{\text e}(t',\theta)$ is the alignment-dependent ionization
probability derived from Eq.~(\ref{free-electron}). For randomly
distributed molecules, $\rho(\theta)$ is a constant. Once the
averaged induced dipoles   $D^{\text {avg}}(\omega)$ are obtained
for a number of different laser intensities, they are then fed into
the propagation equations for the harmonics. The propagation is then
carried out similar to that for atomic targets. We comment that in
this model, dielectric properties of molecules due to non-isotropic
distributions have been neglected.

\subsection{Macroscopic wave packet (MWP)}
As presented in Ref. \cite{jin-2009} [see their Eq.~(25)], the macroscopic HHG
spectrum in the near field for atomic targets can be expressed as
\begin{eqnarray}
\label{hhg-macro}S_{\text
h}(\omega)\propto\omega^{4}|W^{\prime}(\omega)|^{2}|d(\omega)|^{2},
\end{eqnarray}
where $W^{\prime}(\omega)$ is called a ``macroscopic wave packet"
(MWP), and $d(\omega)$ is the PR transition dipole moment. This
relation still holds in the far field [see Eq.~(\ref{total-hhg})],
since the PR transition dipole can be factorized out in
Eq.~(\ref{far-hhg}). The propagation of harmonics in free space to
the far field would thus only modify the MWP.

If molecules are only partially aligned, we calculate $d(\omega)$
by coherently averaging the PR transition dipole weighted by the
ionization probability of $N(\theta)$
\cite{tong-pra-2002,zhao-pra-10}:
\begin{eqnarray}
\label{dip-avg}d^{\text
{avg}}(\omega)=\int^{\pi}_{0}N(\theta)^{1/2}d(\omega,\theta)\rho(\theta)\sin\theta
d\theta.
\end{eqnarray}
From Eq.~(\ref{hhg-macro}), the target structure is reflected in the PR
transition dipole, the propagation effect  of the harmonics, in the meanwhile,
is incorporated in the MWP. The two properties are well separated. The MWP
represents the cumulative effect of the returning electron wave packet (or
microscopic wave packet) after propagation in the medium and in the free space.
The validity of Eq.~(\ref{hhg-macro}) forms the basis of extracting target
molecular structure information from the experimentally measured HHG spectra.

\section{Results and discussion}
\subsection{Macroscopic HHG spectra of Ar: theory vs experiment}

\begin{figure}
\mbox{\rotatebox{270}{\myscaleboxa{
\includegraphics{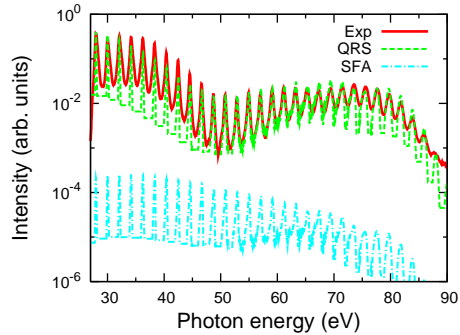}}}}
\caption{(Color online) Comparison of theoretical (QRS: dotted line;
SFA: dot-dashed line) and experimental (solid line) HHG yields of Ar
under a 1200 nm laser. Experimental data are from Ref.
\cite{jin-prl-10}. Laser parameters are given in the text.
\label{Fig1}}
\end{figure}

In Fig.~\ref{Fig1}, we show the macroscopic HHG spectra generated by
a 1200 nm laser. Experimentally \cite{jin-prl-10}, a 0.5 mm-long gas
jet was placed 3 mm after the laser focus (z=3 mm). A vertical slit
with a diameter of 100 $\mu$m is placed 24 cm after the gas jet. The
beam waist at the laser focus is 47.5 $\mu$m, and the laser pulse
duration is $\sim$40 fs. Laser peak intensity in the center of gas
jet was 1.6$\times$10$^{14}~$W/cm$^{2}$, and the pressure of gas jet
was estimated to be 28 Torr. In the simulation, the laser peak
intensity and the pressure are adjusted until the best agreement
with the experiment is reached visually. Using peak intensity of
1.5$\times$10$^{14}~$W/cm$^{2}$, we find the best agreement with
cutoff position in the HHG spectra, and at pressure of 84 Torr, we
find the best agreement in the widths of the harmonics
\cite{jin-prl-10}.

We first calculate the single-atom response either by QRS or SFA. In
SFA, Ar is treated as an effective ``hydrogenlike" atom where the
nuclear charge is chosen such that its 1$s$ binding energy is the
same as the binding energy of Ar. The induced dipole moment is
calculated by using the Lewenstein model \cite{Lewen}. In the QRS,
the ``exact" transition dipole is calculated by using the model
potential given by M\"uller \cite {Muller}. The single-atom response
is then fed into Eq.~(\ref{harm-final}), and the harmonic signals
are collected in the far field in terms of Eq.~(\ref{total-hhg}).

In Fig.~\ref{Fig1}, we can see very good agreement between QRS and
experiment over the photon-energy region of 30-90 eV. The ``famous"
Cooper minimum in Ar \cite{Cooper-min}, is clearly seen  in both
experimental and theoretical spectra. The Cooper minimum in Ar has
been reported in early HHG spectra generated by 800 nm lasers
\cite{hans-prl-09,Mine-pra-08}, and it is seen more  prominently
using long-wavelength lasers \cite{Col-NatPhys-08}. Meanwhile, the
propagated spectra obtained from SFA do not show Cooper minimum, nor
does it reproduce the general spectral shape. Note that Cooper
minimum shall occur in the single-atom HHG spectra but it does not
always appear in the macroscopic HHG spectra. As illustrated in
recent simulations \cite{jin-prl-10}, the position of Cooper minimum
can change or even disappear under different experimental
conditions. Such conclusions are consistent with experimental data
where Cooper minimum may disappear in the HHG spectra by changing
the gas pressure \cite{Mine-pra-08}, or by changing the gas-jet
position with respect to the laser focus
\cite{standford-unpublished}. In the following, we will show that
these changes are due to variations in the MWP, and the separability
of Eq.~(\ref{hhg-macro}) is still valid.

In Fig.~\ref{Fig1}, there are still small discrepancies between the
experimental data and the simulation by QRS despite of various attempts using
somewhat different laser parameters. The harmonic width (or harmonic chirp) in
the simulation is narrower than that in the experimental data. The harmonic
width is mainly determined by laser intensity, pulse duration, and gas pressure
\cite{Var-jmo-05,Zair-prl-08,he-pra-09}. In the experiments, parameters like
pressure of the gas jet, laser intensity and its spatial distribution cannot be
measured precisely. Other factors, like the use of the slit and the position of
the detector can also influence the HHG spectra. All of these uncertainties can
contribute to the discrepancy between the simulation and the measured HHG
spectra. On the other hand, as seen from Fig.~\ref{Fig1}, the overall agreement
between the experiment and the simulation over the 30-90 eV region is quite
satisfactory.

\subsection{Harmonic chirp}

\begin{figure}
\mbox{\rotatebox{0}{\myscaleboxb{
\includegraphics{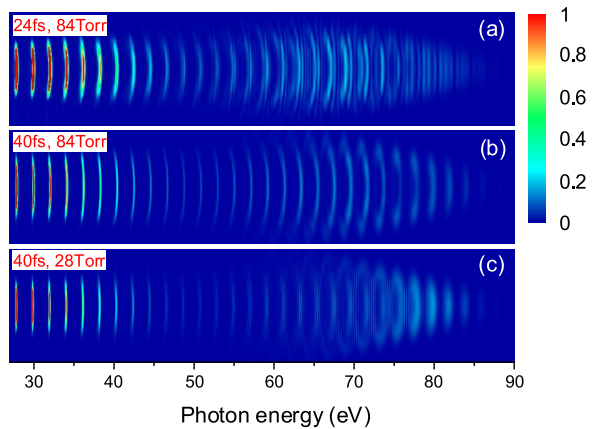}}}}
\caption{(Color online) Spatial distributions of harmonic emission
vs photon energy (normalized using on-axis intensity at 77 eV) in
the far field for the laser with different pulse durations and the
gas jet with different pressures: (a) 24 fs, 84 Torr, (b) 40 fs, 84
Torr, and (c) 40 fs, 28 Torr. The other laser parameters are not
varied and are given in the text. \label{Fig2}}
\end{figure}

Although Fig.~\ref{Fig1} shows the general global agreement between
simulation and experimental data, it is pertinent to examine typical
individual harmonics more carefully. How does the spectral width (or
the harmonic chirp) vary as the pulse duration, gas pressure and
laser intensity are changed? In Fig.~\ref{Fig1} the harmonics were
taken for the gas jet placed after the laser focus, thus short
trajectories were selected. In this case, the harmonic chirp is less
dependent of the laser intensity, especially for a long-wavelength
laser \cite{Ben-oe-06}. We actually vary the laser intensity by
20\%, the harmonic width almost does not change (not shown). We will
concentrate on the effects of the pulse duration and gas pressure
here only.

In Figs.~\ref{Fig2}(a)-\ref{Fig2}(c), we show the spatial
distribution of harmonic emission in the far field (24 cm after the
gas jet) by varying laser duration and gas pressure. All the other
parameters are kept the same as in Fig.~\ref{Fig1}. For each
harmonic, the distribution on the vertical plane is shown.
Integration of harmonic intensity over the vertical dimension in
Fig.~\ref{Fig2}(b) gives the simulated HHG spectra by QRS in
Fig.~\ref{Fig1}. A longer pulse duration and/or a lower pressure
tend to generate sharper (narrower width) lower-order harmonics. A
careful inspection reveals that the peak position of the harmonic
actually blue-shifted from one frame to another. The shift is due to
the change of the fundamental pulse as it propagates through the
nonlinear medium \cite{mett-pra-06}. In addition,   the higher
harmonics are less sharp, reflecting that the quality of phase
matching varies with harmonic orders.

\begin{figure}
\mbox{\rotatebox{270}{\myscaleboxc{
\includegraphics{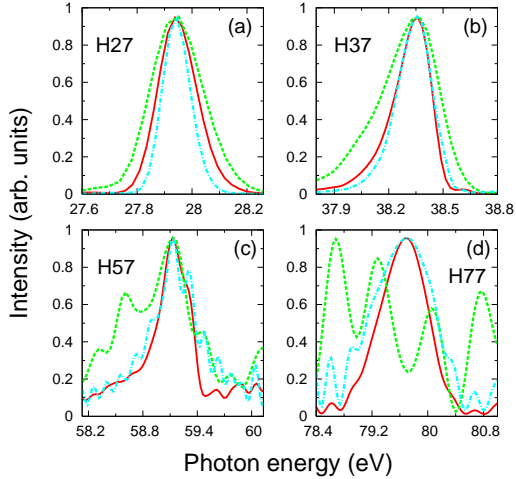}}}}
\caption{(Color online) The dependence of spectral distribution of individual
harmonics on experimental conditions. The spectra are integrated over the
vertical dimension as shown in Fig.~\ref{Fig2}.  Experimental conditions are:
40 fs, 84 Torr (red, solid lines); 24 fs, 84 Torr (green, dashed lines); and 40
fs, 28 Torr (blue, dot-dashed lines). Note that the peak position of the
harmonic in each frame has been shifted to coincide for easy comparison.
\label{Fig3}}
\end{figure}

In Figs.~\ref{Fig3}(a)-\ref{Fig3}(d), we show the spectral
distributions of harmonics H27 (27th harmonic), H37, H57, and H77,
respectively, after they have been integrated along the vertical
dimension. Here we examine the change of harmonic width as the pulse
length and/or gas pressure are varied. Recall that the harmonic
peaks are blue-shifted differently for different conditions. For
easy comparison, the peak position is taken to be from the 40 fs, 84
Torr set (red, solid lines). The spectra from the other two sets are
shifted to have the same peak position. From the figures, for H27
and H37, clearly the harmonic width increases with decreasing pulse
duration. For a given pulse duration, the width increases with gas
pressure. These figures also show that phase matching is not good
for the higher harmonics, especially for the short-duration pulses
where high-energy photons are emitted only from a few half-cycles.
For these higher harmonics, narrower width seems to be obtained by
raising the gas pressure.

The harmonic chirp is a direct consequence of temporal variation of
laser intensity, which can be measured by XFROG (cross-correlation
frequency resolved optical gating) \cite{norin-prl-02,sek-prl-02}.
It is determined by $\Delta\omega(t)=-\partial
\Phi(\omega,t)/\partial t$. The harmonic phase $\Phi(\omega,t)$ is
proportional to laser intensity $I(t)$, with larger proportional
constant for electrons taking long trajectories than for short
trajectories \cite{mett-pra-99} [see their Eq. (1)]. For the
focusing conditions in the present case, only short trajectories are
selected and longer pulse leads to narrower harmonic width. In our
model, Kerr effect on the fundamental field and plasma effect due to
free electron are included in Eq.~(\ref{eff}). The $\Phi(\omega,t)$
is dependent on the gas pressure through $\eta_{2}$ and $n_{\text
e}(t)$. It can be understood that the only variation of gas pressure
could lead to the change of harmonic width. Similar analysis of the
dependence of phase-matching on gas pressure can be found in Refs.
\cite{Pop-pnas-09,Win-rev-08}.

\subsection{Harmonic divergence}
\begin{figure}
\mbox{\rotatebox{270}{\myscaleboxc{
\includegraphics{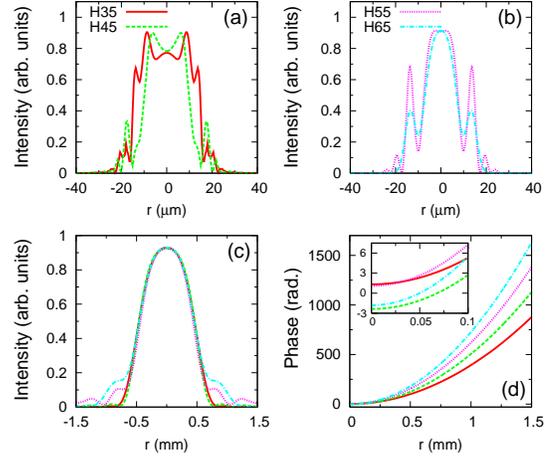}}}}
\caption{(Color online) Spatial dependence of harmonic emission in
the near field (at the exit of gas jet): (a) H35 and H45, (b) H55
and H65, and in the far field (at the entrance of slit): (c)
H35-H65. (d) Spatial dependence of phases for H35-H65 in the far
field. Inset: Phases of H35-H65 in the region close to the
propagation axis. \label{Fig4}}
\end{figure}

We next examine the harmonic emission in the near field and in the
far field. The laser parameters are the same as those in the
simulation in Fig.~\ref{Fig1}. Figs.~\ref{Fig4}(a) and~\ref{Fig4}(b)
show that harmonic emissions in the near field are quite messy
spatially \cite{jin-2009}. Because the harmonics are generated from
a nonuniform Gaussian beam, phase-matching condition in the medium
varies spatially. The radial variation of the phase $\Phi(\omega,r)$
introduces a curvature of the phase front, which makes the harmonic
emission divergent \cite{mett-pra-99}. After further propagating to
the far field, the harmonic emissions become regular, see
Fig.~\ref{Fig4}(c), where harmonics display Gaussian distributions
centered at the propagation axis. In Fig.~\ref{Fig4}(c), only the
short trajectories are selected. The harmonic emission is mainly on
axis because of the small divergence. Since long-wavelength laser is
used \cite{Dou-prl-09}, the divergence in the region from H35 to H65
does not change too much. Similar study of the divergence  of
harmonics has been done for short-wavelength lasers
\cite{Sal-jpb-96}.

We also study the phase of harmonics vs the radial distance in
Fig.~\ref{Fig4}(d). For each harmonic, we note that the calculated
phase grows quadratically with the radial distance and scales almost
linearly with the frequency of the harmonics. The phases near the
propagation axis vary for different harmonics, see the inset of
Fig.~\ref{Fig4}(d). Below we show that the behaviors of intensity
and phase of harmonics in the far field display good (laser-like)
spatial Gaussian character.

Recall that an incident Gaussian beam focused at z=0 propagating
along the z axis \cite{jin-2009,far-field,L'Huillier-1992} is given
by
\begin{eqnarray}
E(r,z)=\frac{bE_{0}}{b+2iz}\exp(-\frac{kr^{2}}{b+2iz})
=|E(r,z)|e^{i\varphi(r,z)} ,
\end{eqnarray}
where
\begin{eqnarray}
\label{gau-int}|E(r,z)|=\frac{bE_{0}}{(b^{2}+4z^{2})^{1/2}}\exp(-\frac{kr^{2}b}{b^{2}+4z^{2}})
,
\end{eqnarray}
and
\begin{eqnarray}
\label{gau-pha}\varphi(r,z)=-\tan^{-1}(\frac{2z}{b})+\frac{2kr^{2}z}{b^{2}+4z^{2}} .
\end{eqnarray}
Here $E_{0}$ is the peak field at the focus, $k$ is the wave vector,
$b=2\pi w_{0}/\lambda$ is the confocal parameter, $w_{0}$ is the
beam waist at the focus, and $\lambda$ is the wavelength. The
intensity of each harmonic  in Fig.~\ref{Fig4}(c) follows the square
of Eq.~(\ref{gau-int}). The phase increases quadratically with r and
linearly with the harmonic order, as seen in Fig.~\ref{Fig4}(d), can
be seen to follow the second term of Eq.~(\ref{gau-pha}). Near r=0,
the phase from the first term of Eq.~(\ref{gau-pha}) also
contributes. This term (multiplied by harmonic order) gives a phase
between -$\pi$ and $\pi$ for each harmonic, as seen in the inset of
Fig.~\ref{Fig4}(d), in which the phase at r=0 is only defined within
2$\pi$. Since for different harmonics the confocal parameters are
probably changed either due to beam waist or wavelength, and the
focused position may also change, we can only claim that each
harmonic beam after propagation is close to a Gaussian beam
qualitatively, but not necessarily quantitatively. Similar study of
the phase of the harmonics in the near field was presented in Ref.
\cite{p.sal-adv}.

\subsection{Harmonic phase vs photon energy in the far field}

\begin{figure*}
\mbox{\rotatebox{270}{\myscaleboxa{
\includegraphics{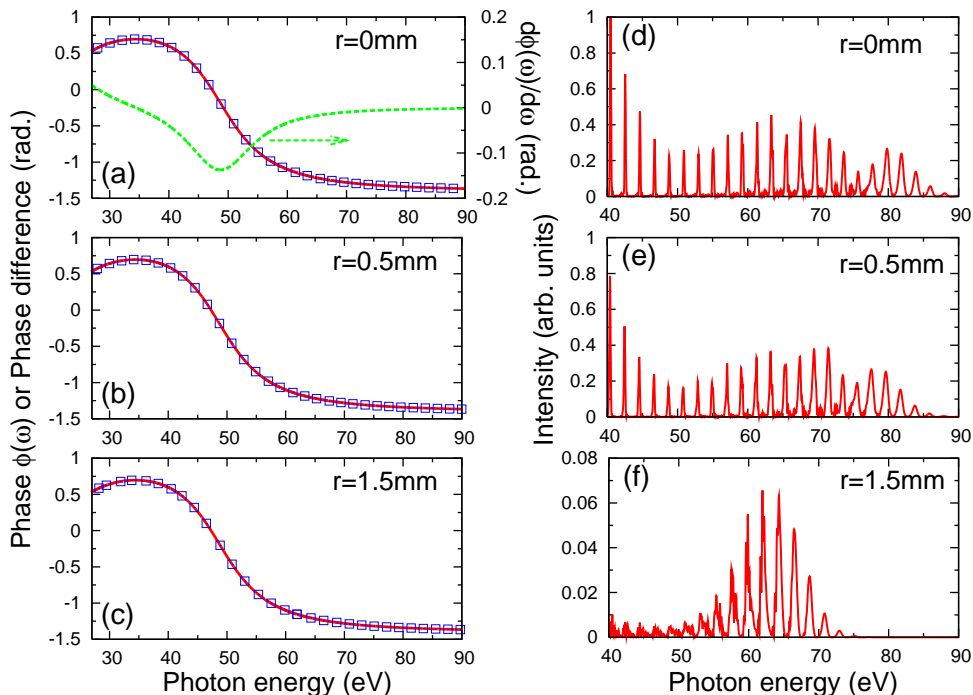}}}}
\caption{(Color online) Phase difference (open squares) between
far-field harmonics by QRS and by SFA, in comparison with the phase
$\phi(\omega)$ (solid lines) of ``exact" transition dipole, at (a)
r=0 mm, (b) r=0.5 mm, and (c) r=1.5 mm. The difference of atto-chip
(the phase difference divided by the energy difference between two
neighboring harmonics) between the QRS and the SFA is shown in (a)
(dashed line). The far-field harmonic emissions at (d) r=0 mm, (e)
r=0.5 mm, and (f) r=1.5 mm vs photon energy. Note that the phases in
(a)-(c) are only taken at harmonic peaks, which are shown in
(d)-(f), respectively. \label{Fig5}}
\end{figure*}

In the QRS model for single-atom response, the phase of each
harmonic is the sum of the phase from the microscopic wave packet
and the phase from the transition dipole [see
Eq.~(\ref{qrs-phase})]. For the harmonics calculated from the SFA,
it is the same sum except that the phase of the transition dipole is
either 0 or $\pi$. Thus the difference in the harmonic phase
calculated from the QRS and from the SFA is given by the phase of
the transition dipole calculated within the QRS (modulus $\pi$). Is
this relation still correct after the propagation, as implied by
Eq.~(\ref{hhg-macro})? In Figs.~\ref{Fig5}(a)-\ref{Fig5}(c), we show
that phase differences (the squares) calculated by QRS and by SFA
for far-field harmonics, and compare them with the phase
$\phi(\omega)$ of the transition dipole from the QRS (the solid
line), at r=0 mm, 0.5 mm, and 1.5 mm, respectively. The laser
parameters are the same as those in Fig.~\ref{Fig1}. The two are in
good agreement. This agreement implies that the phase of the
macroscopic wave packet $W^{\prime}(x,y,\omega)$ obtained from QRS
and from SFA remains identical after propagation at any points
$(x,y)$ on the plane. (In next subsection we will show that the
magnitude of the MWP is also the same.)

For the generation of attosecond pulses, the phase difference (or
atto-chirp) between consecutive harmonics is crucial
\cite{Mar-prl-05}. The above results indicate that atto-chirp
calculated using QRS and SFA differs only by the difference of the
phase $\phi(\omega)$ of the transition dipole between two
neighboring harmonics. This difference divided by 2$\omega_{0}$ (in
units of eV), or the derivative of $\phi(\omega)$ with respect to
$\omega$ (in units of eV) is shown (dashed line) in
Fig.~\ref{Fig5}(a). It is clear that correction to the atto-chirp
calculated from SFA is small, except in the region near the Cooper
minimum where the phase $\phi(\omega)$ changes rapidly. This result
is very significant since it explains why the generation of
attosecond pulses can be understood mostly based on the SFA theory,
even though it does not predict the harmonic spectra accurately. The
phase of each harmonic is mostly determined by the phase of the
returning electron wave packet, which has been accurately accounted
for by the SFA, with very small corrections from the recombination
process. This simplification explains why it is possible to study
the generation of attosecond pulses in the last decade without a
quantitatively accurate theory of HHG.

To appreciate the complexities of harmonics, in
Figs.~\ref{Fig5}(d)-\ref{Fig5}(f) we show the far-field harmonic
emissions at different radial distances $r$ from the propagation
axis. The harmonic emissions for different photon energies are
comparable close to the axis, to provide a broad  energy region from
the on-axis area to synthesize attosecond pulses \cite{Ant-prl-96}.
In the experiment this is accomplished by employing an iris
\cite{Mar-prl-05}. We note that at r=0 mm and r=0.5 mm, the harmonic
spectra resemble each other and the broad Cooper minimum can be
easily identified. This is not the case if the spectra are taken at
r=1.5 mm where the signals are much weaker and the Cooper minimum is
no longer visible. As mentioned in Sec. III A, the ``disappearance"
of Cooper minimum is attributed to the change of MWP, not due to the
recombination process. Experimentally the ``disappearance" of Cooper
minimum in the HHG spectra of Ar has been shown using 800 nm pulses
by changing the experimental conditions
\cite{Mine-pra-08,standford-unpublished}.

\subsection{The dependence of MWP on experimental conditions}
\begin{figure*}
\mbox{\rotatebox{270}{\myscaleboxa{
\includegraphics{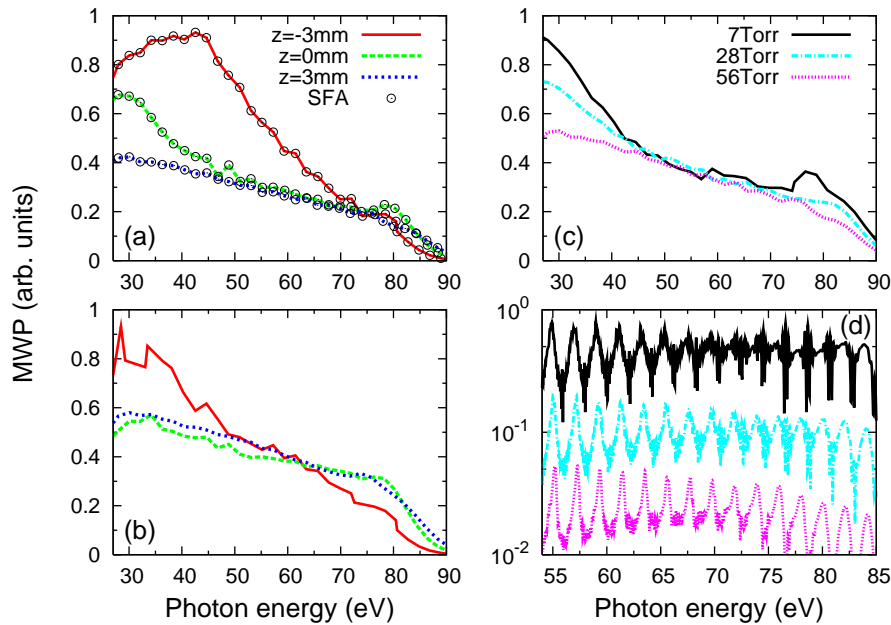}}}}
\caption{(Color online) (a) The magnitude (not the intensity) of
macroscopic wave packets (MWP) at different gas-jet positions: z=-3
mm (before the focus, solid line), z=0 mm (at the focus, dashed
line), and z=3 mm (after the focus, dot-dashed line) by QRS. The
corresponding ones by SFA are also shown (open circles). The HHG
signals are collected at the exit of the slit. (b) Same MWP as in
(a), but the total HHG signals are collected without the slit. The
curves in (a) and (b) have been normalized separately. (c)
Dependence of MWP on gas pressure. Each curve has been normalized by
the pressure. (d) The detailed structure of MWP in (c) from 55 to 85
eV. The curves have been shifted for easy visualization.
\label{Fig6}}
\end{figure*}

In Fig.~\ref{Fig6}(a) we show how the MWP (complex amplitude)
changes with the gas-jet position calculated by the QRS, for HHG
signal collected after the slit. For simplicity, only the envelope
of the MWP is plotted. We keep laser intensity at
1.6$\times$10$^{14}~$W/cm$^{2}$ and gas pressure at 56 Torr. The
slit is placed 24 cm after the gas jet, and all other parameters are
the same as given in Fig.~\ref{Fig1}. We also show the MWP
calculated by SFA under exactly the same conditions, and the results
are shown (in circles) in Fig.~\ref{Fig6}(a). The comparison shows
that the MWPs (magnitude) from QRS and from SFA are the same for the
same experimental condition, even though the MWP can change greatly
depending on the gas-jet position. When the gas jet is placed
``after" the laser focus (z=3 mm) the MWP is very flat, since good
phase-matching is favored for this arrangement as the single-atom
harmonic phase is partially cancelled by the Gouy phase from the
focused laser \cite{p.sal-prl-1998,p.sal-adv}. If the gas jet is
placed before the laser focus, we note that the MWP changes rapidly,
especially near photon energy around 50 eV. Such strong energy
dependence can wash out the Cooper minimum in the HHG spectra
\cite{jin-prl-10}.

In Fig.~\ref{Fig6}(b) we compare the MWPs as in Fig.~\ref{Fig6}(a)
but without the slit. In such comparison, the harmonics from
electrons following long and short trajectories are both collected.
By comparing Figs.~\ref{Fig6}(a) and \ref{Fig6}(b), we note that the
MWPs for gas jet at z=-3 mm and z=0 mm change significantly, but for
z=3 mm, the MWPs in both cases remain rather flat. This shows that a
narrow slit in the far field can select the short trajectories
effectively.

In Fig.~\ref{Fig6}(c) we investigate how the MWP depends on the gas
pressure for the focusing condition of z=3 mm. The MWP has been
normalized by the ratio of the pressure. The three curves would be
on top of each other if a complete phase-matching condition had been
fullfilled. The curve for higher pressure is slightly lower
indicates that full phase matching is not reached, especially for
the lower harmonics. With the increase of pressure, the MWP is much
smoother vs energy. In fact, increasing the gas pressure tends to
smooth out the harmonics. These results also indicate that the
harmonic energy increases quadratically with the gas pressure, which
is in agreement with measurements reported in Ref.
\cite{Shiner-prl-2009}. In Fig.~\ref{Fig6}(d) we display the full
MWP vs the gas pressure. It shows that the  harmonics exhibit better
shaped peaks  as the phase-matching condition is favored at higher
pressure \cite{Pop-pnas-09,Win-rev-08}.

\subsection{Wavelength scaling of harmonic efficiency}
\begin{figure*}
\mbox{\rotatebox{270}{\myscaleboxa{
\includegraphics{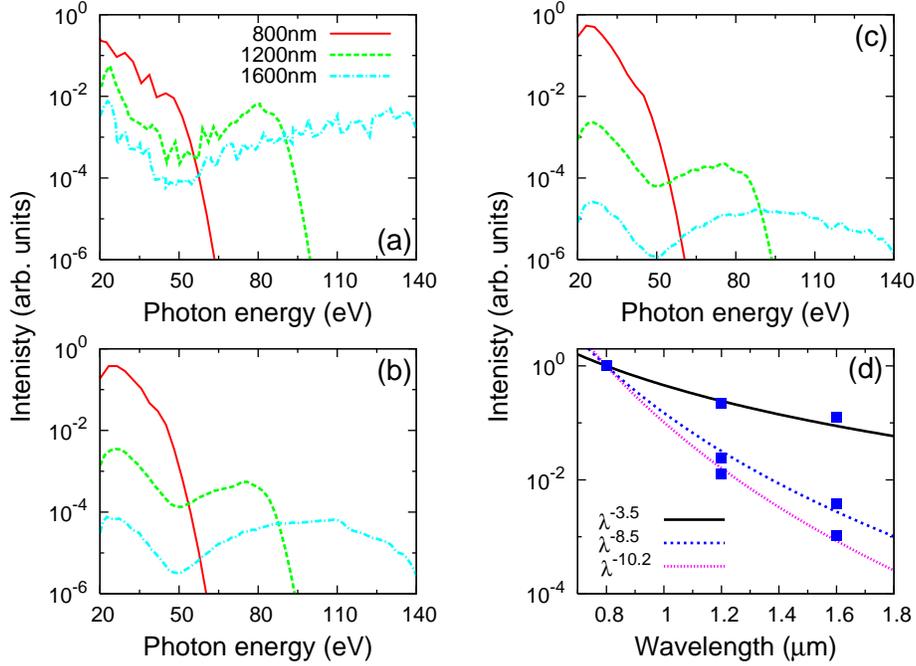}}}}
\caption{(Color online) (a) Single-atom HHG spectra, and macroscopic
HHG spectra  without (b) and  with (c) the slit for 800 nm (solid
lines), 1200 nm (dashed lines) and 1600 nm (dot-dashed lines)
lasers. The laser parameters are given in the text. (d) The
wavelength dependence of the total integrated HHG yields above 20
eV. The integrated HHG yields in (a), (b), and (c) follow
$\lambda^{-3.5\pm 0.5}$, $\lambda^{-8.5\pm 0.5}$, and
$\lambda^{-10.2\pm 0.2}$, respectively. \label{Fig7}}
\end{figure*}

One of the main interests in the study of HHG is to produce bright
tabletop XUV or soft X-ray sources, or intense attosecond pulses.
Since the single-atom harmonic cutoff energy is proportional to the
square of the wavelength of the driving laser, HHG generated by
mid-infrared (MIR) lasers has been of great interest experimentally.
While MIR lasers can efficiently reach high-energy photons, the
yield is less favorable. It is of interest to study how the HHG
yield scales with the laser wavelength. Within the single-atom
response level, there have been a few theoretical calculations
\cite{Tate-prl-2007,Sch-prl-07,Frolov-prl-08,Gordon-oe-05,Lan-pra-2010}.
However, to compare with experimental HHG spectra, macroscopic
propagation effect has to be included. A few investigations on the
wavelength scaling of HHG experimentally
\cite{Shiner-prl-2009,Col-NatPhys-08,Yak-oe-07,Pop-ol-08,fal-apl-2010}
have been reported. However, theoretical analysis is still rather
scarce.

To study wavelength scaling of the HHG yields, one has to fix all
other parameters that may affect the efficiency of HHG. One also has
to decide if it is the total HHG yield or only the HHG yield within
a given photon energy region. In single-atom simulations, the laser
parameters can be easily fixed. But this is not the case in
experiments. Theoretical simulations including macroscopic
propagation effect are few \cite{fal-oe-09}. Since the resulting HHG
spectra depend on so many other parameters, as we have demonstrated
in the earlier subsections, any wavelength scaling laws derived are
likely to depend on experimental parameters used. In spite of this
limitation, it is still of interest to take a look at the wavelength
scaling by using the present QRS model. For this purpose, we will
define a parameter that describes the efficiency of harmonic
generation. This is the ratio between the output energy (total
harmonic energy) with respect to the input energy (fundamental laser
energy) for different laser wavelengths. According to Ref.
\cite{Tosa-pra-2003}, the input energy E$_{\text {pulse}}$ can be
related to the laser duration $\tau_{\text p}$ and peak intensity
I$_{0}$ at the focus by $E_{\text {pulse}}=\frac{I_{0}}{2}\pi
w_{0}^{2}\tau_{\text p}$ if the laser beam has a Gaussian
distribution in time and space. The output energy can be obtained by
integrating the harmonic intensity over a photon-energy region:
\begin{eqnarray}
E_{\text {out}}=\int_{\omega_{\text {min}}}^{\omega_{\text
{max}}}\int \int|E_{\text h}(x,y,\omega)|^{2}dx dy d\omega.
\end{eqnarray}.

In Fig.~\ref{Fig7}(a) we show the single-atom HHG spectra calculated
for three wavelengths. Only the envelope of each spectra is shown.
In the calculation, the laser intensity and duration are kept as
1.6$\times$10$^{14}~$W/cm$^{2}$ and 40 fs, respectively. In
Fig.~\ref{Fig7}(b) the HHG spectra obtained after including
macroscopic propagation are shown. In the calculation, the beam
waist at the focus is kept as 47.5 $\mu$m, a 0.5 mm-long gas jet is
placed at 3 mm after the laser focus, and gas pressure is kept at 56
Torr. The yield of each harmonic is obtained by integrating over the
whole plane perpendicular to the propagation axis. In
Fig.~\ref{Fig7}(c), the HHG yields are recorded after they have
passed a slit (the slit with a diameter of 100 $\mu$m is placed at
24 cm after the gas jet). From Figs~\ref{Fig7}(b) and \ref{Fig7}(c),
we calculate the HHG efficiencies per atom vs the wavelength.

In Fig.~\ref{Fig7}(a), the ratio of input energy is 1:1:1 for 800
nm, 1200 nm, and 1600 nm lasers. If we integrate the HHG yields
above 20 eV as the output energy. The resulting energy follows
$\lambda^{-3.5\pm0.5}$ shown in Fig.~\ref{Fig7}(d). If we integrate
the HHG yields between 20 eV and 50 eV, which would give a scaling
rule of $\lambda^{-5}$. In Tate {\it et al}. \cite{Tate-prl-2007},
the laser intensity and the number of optical cycles were fixed for
800 nm and 2000 nm lasers. According to our approach, the ratio of
input energy is 1:2.5 for the 800 nm and 2000 nm lasers. And their
scaling rules at constant intensity of $\lambda^{-(5-6)}$ would be
modified as $\lambda^{-(6-7)}$ at a constant input energy.

When propagation effect is considered it is generally
known~\cite{Tate-prl-2007,Shan-pra-01} that phase-matching condition
is more difficult to meet for longer wavelength, thus the HHG
efficiency decreases with increasing wavelength. Here we consider
the total HHG yields for the lasers used in Fig.~\ref{Fig7}(b) in
which the gas jet is placed at z=3 mm. Since the laser intensity is
fixed at the center of the thin gas jet, we calculate that the
intensities at the laser focus are 1.78$\times$10$^{14}~$W/cm$^{2}$,
2.01$\times$10$^{14}~$W/cm$^{2}$, and
2.33$\times$10$^{14}~$W/cm$^{2}$, for 800 nm, 1200 nm, and 1600 nm
lasers, respectively, thus the input energies have the ratios of
1:1.13:1.31. We find that HHG yields integrated from 20 eV up scale
like $\lambda^{-8.5\pm0.5}$, as shown in Fig.~\ref{Fig7}(d). If we
only integrate the harmonics between 20-50 eV, then the scaling rule
is $\lambda^{-10.5}$.

Experimentally, Colosimo {\it et al}. \cite{Col-NatPhys-08} reported
that the HHG yields between 35-50 eV for 2000 nm lasers are about
1000 times smaller than that for 800 nm lasers, for experimental
conditions that were kept ``as fixed as possible". This would give a
$\lambda^{-9}$ dependence in this narrow energy region which is not
too far off from our scaling of $\lambda^{-10.5}$. In addition,
Shiner {\it et al}. \cite{Shiner-prl-2009} reported a scaling rule
of $\lambda^{-6.3\pm1.1}$ for the HHG of Xe with a fixed laser
intensity. By assuming  perfect phase-matching condition for all the
laser wavelengths used, they derived scaling law of
$\lambda^{-6.3\pm1.1}$ that was to be compared to the scaling law
derived from the single-atom response. We cannot compare their
results with our simulation. They also used a Bessel beam (instead
of Gaussian beam) in the experiment and the gas jet was located at
the laser focus. Since the HHG yields depend on so many experimental
parameters, it is clear that any simple scaling laws derived should
be taken with caution. In Fig.~\ref{Fig7}(d), we also show the
scaling law for the case where the HHG yields are collected after
the slit. We integrate the HHG signals above 20 eV and obtain the
$\lambda^{-10.2\pm0.2}$ scaling. In general, good phase-matching
condition becomes more difficult to meet with increasing laser
wavelength. Even if the gas jet is placed after the laser focus, the
short trajectories are not selected efficiently for longer
wavelength lasers. A slit is usually used to select contributions
from short trajectories in the far field. By blocking out
contributions [see Fig.~\ref{Fig7}(c)] from the long trajectories
the harmonic efficiency becomes worse.

Based on the above analysis, the HHG yields for long-wave driving
lasers under the same experimental conditions appear quite
unfavorable. On the other hand, for practical purpose,
experimentally high harmonics are to be generated with optimized
conditions. In Colosimo {\it et al}. \cite{Col-NatPhys-08} it was
reported that the HHG yields between 35-50 eV generated by using
2000 nm lasers can be as high as 50$\%$ of that from 800 nm lasers
if the experimental conditions were optimized independently.
Furthermore, Chen {\it et al}. \cite{chen-jila-10} demonstrated that
it was possible to use much higher pressure to generate HHG for long
wavelength lasers, thus achieving usable photon yields even in the
water-window region. Clearly additional theoretical analysis of
macroscopic propagation effects on HHG for long-wavelength driving
lasers under different experimental conditions is desirable.

\subsection{Macroscopic HHG spectra of N$_{2}$ in an 800 nm laser}
\begin{figure*}
\mbox{\rotatebox{270}{\myscaleboxa{
\includegraphics{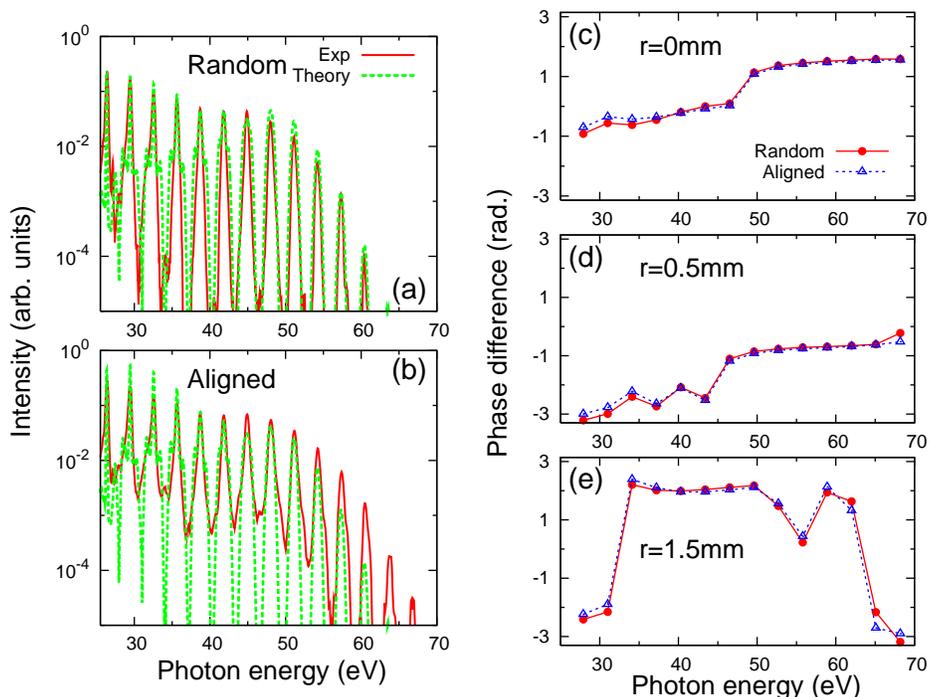}}}}
\caption{(Color online) Comparison of theoretical (dashed lines) and
experimental (solid lines) HHG yields from (a) random and (b)
partially aligned N$_{2}$ molecules in an 800 nm laser. Experimental
data are from Ref. \cite{hans-prl-10}. Phase difference between
consecutive harmonics of random (solid circles) and partially
aligned (open triangles) N$_{2}$ molecules in the far field at
different radial distances: (c) r=0 mm, (d) r=0.5 mm, and (e) r=1.5
mm. \label{Fig8}}
\end{figure*}

The QRS has been used to calculate HHG spectra from single
molecules. In order to compare with experimental data, it is often
assumed that HHG measured in the experiment are taken under the
perfect phase-matching conditions. While such a model has been shown
to be successful in interpreting a number of experimental
observations \cite{at-2009,at-prl-2009,at-jpb-2009}, it is still
crucial to understand the effect of propagation on the HHG in the
medium. In Figs.~\ref{Fig8}(a) and \ref{Fig8}(b) we show the
macroscopic HHG spectra generated by N$_{2}$ molecules that are
isotropically distributed or partially aligned along the
polarization axis of an 800 nm laser. The HHG spectra have been
reported recently \cite{hans-prl-10} using 800 nm and 1200 nm
lasers, and the results from 1200 nm laser have been recently
analyzed \cite{jin-prl-10}. To achieve good agreement with
experiment in the cutoff region, we need to use peak intensity of
1.8$\times$10$^{14}~$W/cm$^{2}$ instead of
2.3$\times$10$^{14}~$W/cm$^{2}$ in the experiment. We use the other
parameters as close as those given in the experiment: pulse duration
is $\sim$ 32 fs, beam waist at the focus is $\sim$ 40 $\mu$m, and
the slit with a diameter of 100 $\mu$m is placed at 24 cm after the
gas jet. A 1 mm-wide gas jet is located 3 mm after the laser focus.
Using the QRS, we first obtain induced dipoles for fixed-in-space
molecules for different laser intensities. They are then averaged
coherently according to the alignment distribution of molecules by
Eq.~(\ref{avg-in-dip}). The resulting induced dipoles are then fed
into Eq.~(\ref{harm-final}). In the experiment, the degree of
alignment was estimated to be $\langle \cos^{2}\theta
\rangle$=0.6-0.65, we use an alignment distribution of
$\cos^{4}\theta$ in the simulation. Note that only HOMO is included
in the calculation. This is adequate since contributions from HOMO-1
is important only for molecules that are nearly perpendicular to the
polarization axis \cite{at-jpb-2009,stanford-Sci-N2}.

\begin{figure}
\mbox{\rotatebox{270}{\myscaleboxa{
\includegraphics{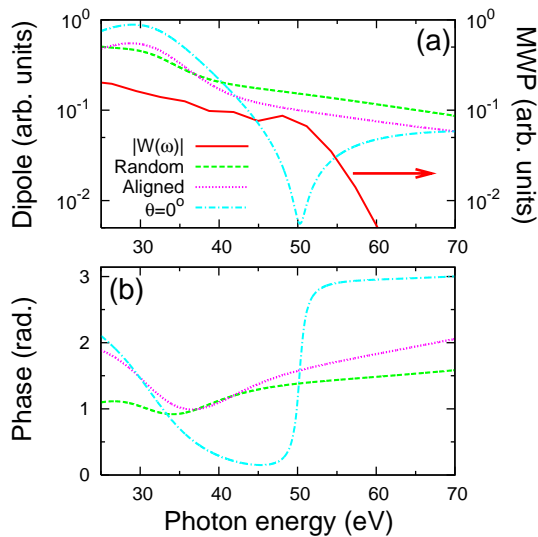}}}}
\caption{(Color online) (a) Macroscopic wave packet (MWP) for an 800
nm laser (solid line), and averaged photoionization (PI) transition
dipole moments (parallel component) for randomly distributed (dashed
line), partially aligned (dotted line), and perfectly aligned
($\theta$=0$^{o}$, dot-dashed line) N$_{2}$ molecules. (b) The
photon-energy dependence of the phase for averaged PI transition
dipoles in (a). \label{Fig9}}
\end{figure}

Figs.~\ref{Fig8}(a) and \ref{Fig8}(b) show the good overall
agreement between experiment and theory for both randomly
distributed and partially aligned N$_{2}$. The experimental spectra
reveal a weak minimum at 39$\pm$2 eV \cite{hans-prl-10} for both
aligned and unaligned molecules. The theory also predicts a minimum
near 45 eV, and the position of minimum is not shifted from random
to the partially aligned N$_{2}$. For aligned N$_{2}$, we note that
McFarland {\it et al}. \cite{McF-pra-09} reported a weak minimum
around 39 eV in the HHG spectra using an 800 nm laser. But the
minimum was observed at about 45 eV in Torres {\it et al}. \cite
{Torres-oe-2010} using an 800 nm laser, and at about 38 eV using a
1300 nm laser for unaligned N$_{2}$ (see their Fig.~3). For 1200 nm
lasers, minimums in HHG spectra from the experiment
\cite{hans-prl-10} and the simulation \cite{jin-prl-10} have also
been reported. The exact location of the minimum is not always
identical since it can be somewhat altered due to the energy
dependence of the MWP [see Fig.~\ref{Fig9}(a)], which changes with
laser parameters and experimental conditions. Despite of the
difference in the positions of the minima between the simulation and
experiment, we consider that the overall agreement in the two
spectra is quite satisfactory.

In Figs.~\ref{Fig8}(c)-\ref{Fig8}(e) we show the phase difference
between neighboring harmonics for randomly distributed and partially
aligned N$_{2}$ molecules in the far field, for different radial
positions from the propagation axis. As mentioned in Sec. III D,
phase differences, which reveal emission times of harmonics
\cite{Mairesse-sci-2003} and can be measured by RABITT
(reconstruction of attosecond beating by interference of two-photon
transition) \cite{paul-sci-01,boutu-NatPhys-2008} technique, are
significant for the generation of attosecond pulses
\cite{Wab-epjd-06}. Near the axis, we note that the phase difference
is linearly changed vs photon energy in the plateau ending up by a
sharp rise near about 48 eV in Fig.~\ref{Fig8}(c) and 45 eV in
Fig.~\ref{Fig8}(d), and beyond the cutoff it is almost a constant.
The nearly constant phase difference was observed by Mairesse {\it
et al}. \cite{mairesse-prl-04} for atomic targets to optimize the
conditions for attosecond pulse generation. Since the phase changes
rapidly near the minimum of the harmonic spectra, this explains that
the position of the HHG minimum in the integrated spectra can be
easily changed, depending on how the integrated spectra are
measured. We comment that the phase difference vs photon energy in
Fig.~\ref{Fig8}(e) is less regular. These harmonics, taken at
position away from the axis, have large contributions from long
trajectories and they are not suitable for attosecond pulse
generation. Finally, phase difference for randomly distributed and
partially aligned N$_{2}$ agree well. The small variation comes from
the slightly different derivative of the phase of averaged
transition dipoles for random and partially aligned N$_{2}$ [see
Fig.~\ref{Fig9}(b)].

To understand the results shown in Fig.~\ref{Fig8}, we show the MWP
for the laser used, and photoionization (PI) transition dipole
moment (parallel component only) for N$_2$ molecules fixed in space,
as well as the averaged PI transition dipoles defined by
Eq.~(\ref{dip-avg}) for randomly distributed and partially aligned
N$_2$ molecules in Fig.~\ref{Fig9}. According to
Eq.~(\ref{hhg-macro}), macroscopic HHG yields can be expressed as
the product of a MWP and an averaged transition dipole. The MWP
(magnitude) is identical for random or partially aligned N$_{2}$
under the same laser and experimental condition. The transition
dipole for fixed-in-space molecules shows a minimum at photon energy
that depends on the angle between the molecular axis and the laser
polarization axis (see Refs. \cite{at-2009,jin-pra-10}). An average
over the angular distribution of the molecules washes out the
minimum, except for a relatively faster drop of the transition
dipole near 35-45 eV. The effect of angular average also washes out
the rapid phase change in the transition dipole moment. For
molecules fixed at $\theta=0^\circ$, Fig.~\ref{Fig9}(b) shows a
rapid phase change of near $\pi$ in a very narrow energy region near
50 eV, i.e., at the position of the minimum in Fig.~\ref{Fig9}(a).
However, after the angular average, one can only see somewhat faster
phase change at small photon energies. Since both the MWP and the
transition dipole exhibit minor energy dependence, the actual
minimum position of the HHG is difficult to locate accurately. With
much better aligned molecules, the position of the minimum can
probably be better determined. According to Eqs.~(\ref{hhg-macro})
and (\ref{dip-avg}), the averaged PI transition dipole can be
obtained from the experimental HHG spectra, and it could be used to
retrieve the alignment-dependent ionization probability,
$N(\theta)$. This may provide another method to check the calculated
$N(\theta)$ \cite{zhao-pra-10}.

\section{Summary and outlook}

In this paper we have described a complete theory for high harmonic
generation (HHG) in a macroscopic atomic or molecular medium. Our
approach is based on the simultaneous solution of the coupled
Maxwell's equations describing macroscopic propagation of both
driving laser pulse and its high harmonic fields together with the
microscopic induced dipoles. For the latter we use the recently
developed quantitative rescattering (QRS) theory for a
single-atom/molecule response. This scheme provides a simple and
efficient method for calculating HHG from a macroscopic medium,
which is otherwise formidable, especially in the case of molecular
targets.

Our results show quantitative good agreements with recent
experimental HHG measurements \cite{jin-prl-10,hans-prl-10} either
for Ar or N$_{2}$ targets. For different laser and experimental
conditions, we present the detailed analysis of HHG intensity and
phase. Since the calculation which includes macroscopic propagation
for an isotropic or an aligned molecular target is quiet scare
\cite{jin-prl-10}, we hope that this paper will further stimulate
the interest in establishing quantitative theory for HHG, which can
compare directly with real experiments for partially aligned media.
We note that in this paper the effect of absorption and dispersion
of high harmonics are neglected for molecular N$_2$, which should be
adequate at low gas pressure. In general, absorption and dispersion
are anisotropic in case of aligned molecules, which are not
available generally in the literature. One of the most important
results of this paper is that at the macroscopic level under typical
experimental conditions HHG spectra can be factorized as a product
of a macroscopic wave packet and a photorecombination transition
dipole. The latter is a property of the target only. This
factorization provides a solid foundation for extracting the
molecular structures from HHG spectra. As demonstrated recently
\cite{hans-nature-10}, HHG has been used for ultrafast probes of
excited molecules, our work also provides the needed theoretical
basis for that.

\section{Acknowledgments}
This work was supported in part by Chemical Sciences, Geosciences
and Biosciences Division, Office of Basic Energy Sciences, Office of
Science, U.S. Department of Energy.

\end{document}